\documentclass[conference]{IEEEtran}
\IEEEoverridecommandlockouts

\usepackage{cite}
\usepackage{amsmath,amssymb,amsfonts}
\usepackage{algorithmic}
\usepackage{graphicx}
\graphicspath{{svg-inkscape/}}
\usepackage{tabularx}
\newcolumntype{Y}{>{\centering\arraybackslash}X}
\usepackage{textcomp}
\usepackage{xcolor}
\usepackage{url}
\usepackage{tikz}
\usetikzlibrary{positioning,arrows.meta,fit,backgrounds,calc}
\def\BibTeX{{\rm B\kern-.05em{\sc i\kern-.025em b}\kern-.08em
    T\kern-.1667em\lower.7ex\hbox{E}\kern-.125emX}}
\begin{document}

\title{Beyond DER: Speaker Counting in Crowded End-to-End Diarization}

\author{
  \IEEEauthorblockN{Lahiru Samarakoon, Hongxu Zhu}
  \IEEEauthorblockA{
    \textit{Fano}, Hong Kong SAR, China \\
    \{lahiru, joe.zhu\}@fano.ai
  }
}


\maketitle

\begin{abstract}
Speaker diarization must solve two problems: counting how many speakers are present
in a given conversation, and assigning speech to each one. This becomes harder in
crowded conversations with five or more speakers. We evaluate how end-to-end neural
diarization models count speakers, and find systematic under-counting in crowded
recordings. The standard diarization error rate (DER) hides this failure, because it
is duration-weighted and barely penalizes the dropped, low-activity speakers. We
therefore also report the Jaccard error rate (JER) and explicit counting metrics. We
propose a gated loss that couples speaker existence with frame activity. This loss
can be computed in two ways, duration-weighted or speaker-weighted, and the
speaker-weighted variant, used as a regularizer, reduces the under-count. On
real-world crowded recordings our method clearly improves the counting metrics,
lowering JER by about 6\% relative and DER by about 9\% relative, while leaving
sparse recordings unharmed.
\end{abstract}


\section{Introduction}

Speaker diarization answers the question of who spoke when, recognizing the
different speakers in a conversation and tracking their participation over time.
This is difficult in practice: the number of speakers is usually unknown
beforehand, and speech frequently overlaps as people interrupt one another
\cite{ryant2020third, chung2020spot, kinoshita21_interspeech}. Traditional methods
cluster speaker embeddings and chain together several components, typically a voice
activity detector, an embedding extractor, and a clustering backend, which
complicates deployment and struggles with overlapping speech because each frame is
assigned to a single speaker \cite{anguera2012speaker, park2022review, diez2018but}.
End-to-end neural diarization (EEND) addresses both issues by formulating
diarization as a per-speaker multi-label classification problem, predicting each
speaker's activity at every frame with one model that handles
overlap \cite{fujita2019end}.

Early EEND models predict a fixed number of speakers, a limitation rooted in the
permutation-invariant training (PIT) used to match predictions to references
\cite{yu2017permutation, fujita19eend}. Attractor-based formulations remove
this restriction: EEND-EDA generates a variable set of speaker attractors with an
LSTM encoder-decoder \cite{horiguchi20EENDEDA, horiguchi22EENDEDA}, and EEND-TA replaces the recurrent
module with Transformer attractors, in line with the broader shift from recurrent
layers to attention
\cite{samarakoon2023transformer, transformer, gulati2020conformer, broughton2023improving, rybicka2022end, fujita2023intermediate}. These attractor models have since been pushed toward larger and unbounded speaker counts \cite{horiguchi21global_local, kanda2022transcribe}.

Despite these advances, both EEND-EDA and EEND-TA systematically under-count as
conversations grow crowded, which we define here as five or more speakers. In these
cases they silently drop the speakers who contribute little speech. This failure is easy to overlook. The dominant metric, the diarization error rate
(DER)~\cite{ryant2020thirdeval}, is duration-weighted, so talkative speakers dominate the score. A
model can therefore report a healthy DER while omitting low-activity speakers that,
from a user's perspective, still matter.
Measuring counting therefore needs metrics that DER alone does not provide, such as
the Jaccard error rate (JER) and explicit counting accuracy \cite{ryant2019second}.
We analyze this under-counting in detail in Section~\ref{sec:analysis}.

At its core, diarization interleaves two questions: how many distinct speakers a
recording contains, and to which of them each speech frame belongs. We argue that this crowded under-counting stems largely from how the models are
trained. The training objective works against both questions. It is \emph{decoupled},
supervising speaker existence and frame activity separately. It is also
\emph{duration-weighted}, so low-activity speakers receive little gradient. These
speakers therefore end up neither reliably counted nor assigned frames. Building on EEND-TA, we propose a training-only remedy that leaves the backbone and
inference unchanged, in the spirit of training-time regularizers for diarization~\cite{yu2022auxiliary, samarakoon2025variance}. The remedy is a gated loss, inspired by set-based detection~\cite{carion2020detr,harkonen2024eend}. It couples existence with activity, so an attractor is credited only when it both exists and is active.
The gated loss can be used alone or as a hybrid regularizer. It can weight every
speaker equally, regardless of airtime, rather than by how much they speak. This
shifts supervision toward the low-activity speakers that crowded recordings tend to
drop. The contributions of this work are as follows:
\begin{itemize}
\item We show that both EEND-EDA and EEND-TA systematically under-count in crowded
recordings, and that the duration-weighted DER masks this failure. We therefore
evaluate counting directly with JER, exact-count accuracy, $\pm 1$ accuracy, and
count bias.
\item We propose a gated loss that couples speaker existence with frame activity,
usable alone or as a hybrid regularizer. We pair it with a speaker-weighted reduction
that supervises every speaker equally, regardless of airtime.
\item We show that the speaker-weighted reduction significantly improves speaker counting on crowded recordings. 
These gains come at no cost to DER, do not harm sparse recordings, and are consistent across domains.
\end{itemize}

\section{Speaker Counting Analysis}
\label{sec:analysis}

DER alone does not capture how well a system counts speakers. We therefore evaluate
counting with several metrics that complement it:
\begin{itemize}
\item Jaccard Error Rate (JER) \cite{ryant2019second}: A speaker-weighted metric that is less forgiving than DER for missing low-activity speakers.
\item \emph{Exact-count accuracy}: The fraction of recordings where the predicted speaker count matches the reference count.
\item \emph{$\pm 1$ accuracy}: Allows for a one-speaker error, recognizing that a near-miss is more forgivable than a gross error.
\item \emph{Count bias}: the mean signed difference between predicted and reference counts, $\mathrm{CB} = \frac{1}{R}\sum_{r=1}^{R}(\hat{s}_r - s_r)$ for $R$ recordings with predicted count $\hat{s}_r$ and reference count $s_r$; a negative value indicates under-counting.
\end{itemize}

We study two attractor-based backbones that differ only in how attractors are generated: 
the autoregressive (AR) EEND-EDA~\cite{horiguchi22EENDEDA} and the non-autoregressive (NAR) EEND-TA~\cite{samarakoon2023transformer}. 
We train both under the same data and training recipe. Any failure they share therefore cannot be attributed to attractor generation. 
We restrict the analysis to recordings with at most eight speakers and evaluate both with the metrics above.

\begin{figure}[t]
  \centering
  \def\svgwidth{\linewidth}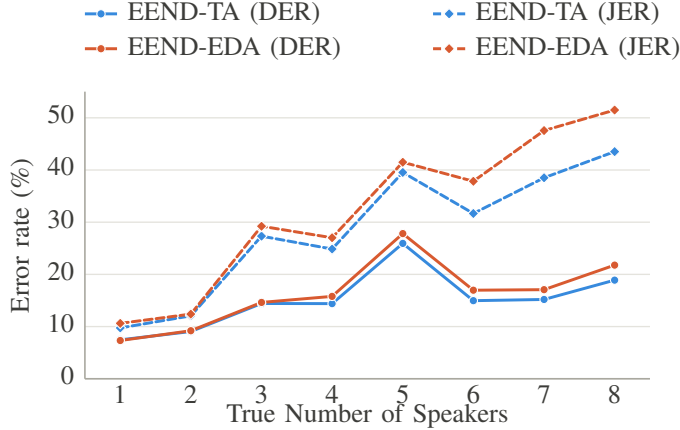
  \caption{Comparison of DER and JER relative to the true number of speakers. 
  Results are pooled across the test sets.}
  \label{fig:der_jer_eda_ta}
\end{figure}

Figure~\ref{fig:der_jer_eda_ta} compares DER and JER as a function of the true speaker count.
For both EEND-EDA and EEND-TA, performance degrades as the number of speakers increases. 
Notably, JER worsens more steeply than DER because its speaker-weighted nature heavily penalizes counting mistakes. 
Across speaker counts, EEND-TA generally achieves lower error rates than EEND-EDA on both metrics, and its advantage tends to be larger in
recordings with more speakers. However, robust speaker counting in crowded recordings remains a critical challenge
for both architectures, regardless of how attractors are generated.




\begin{figure}[t]
\centerline{\def\svgwidth{\linewidth}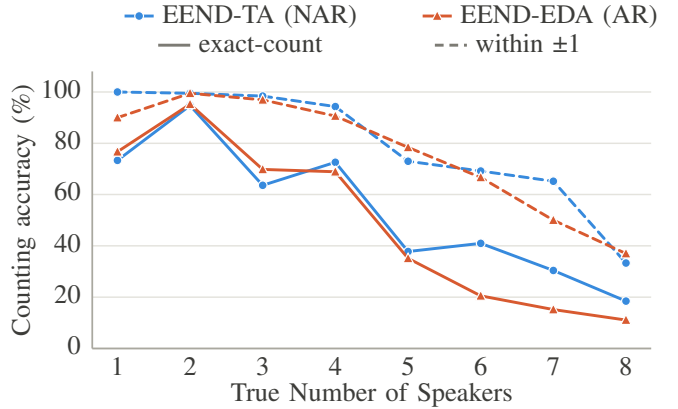}
\caption{Counting accuracy versus the true number of speakers.}
\label{figure_1}
\end{figure}

Figure~\ref{figure_1} presents the two accuracy metrics as a function of the true speaker count. 
Both models count reliably up to roughly four speakers, after which accuracy declines steeply. By eight speakers, exact-count accuracy drops below 20\%. Even the tolerant $\pm 1$ measure falls to approximately 35\%. This decline follows the same trend for both paradigms. 
While EEND-TA is somewhat more robust in the mid-range, neither architecture avoids the eventual collapse. 
That both sequential and parallel attractor generators fail in the same way suggests
the problem is not specific to how attractors are generated, but is shared across
these EEND models.

\begin{table}[t]
\caption{Count bias $\mathrm{CB}$ versus the true number of speakers, pooled over the test sets. A value of $0$ is a perfect count; negative values indicate under-counting.}
\begin{center}
\begin{tabularx}{\columnwidth}{|Y|Y|Y|}
\hline
\textbf{True \#spk} & \textbf{EEND-TA} & \textbf{EEND-EDA} \\
\hline
1 & $+0.27$ & $+0.33$ \\
\hline
2 & $+0.05$ & $+0.04$ \\
\hline
3 & $-0.05$ & $-0.10$ \\
\hline
4 & $-0.04$ & $-0.02$ \\
\hline
5 & $-0.41$ & $-0.35$ \\
\hline
6 & $-0.90$ & $-0.79$ \\
\hline
7 & $-1.11$ & $-1.61$ \\
\hline
8 & $-1.93$ & $-1.89$ \\
\hline\hline
5--8 & $-1.03$ & $-1.13$ \\
\hline
1--8 & $-0.17$ & $-0.19$ \\
\hline
\end{tabularx}
\label{tab:count_bias}
\end{center}
\end{table}

While accuracy metrics show \emph{that} counting fails, Table~\ref{tab:count_bias} reveals \emph{how}. 
Count bias is near zero up to four speakers but turns increasingly negative, reaching $-1.9$ at eight. 
In crowded recordings the models \emph{under}-predict by roughly two speakers, missing low-activity speakers that DER scarcely penalizes. 

This under-count is a failure of both counting and frame assignment. Because EEND-TA generally counts at least as well as EEND-EDA and achieves lower overall DER, we adopt it as the backbone for the remainder of the paper.

\section{Proposed Method}
\label{sec:method}

\begin{figure*}[t]
\centerline{\includegraphics[width=\textwidth]{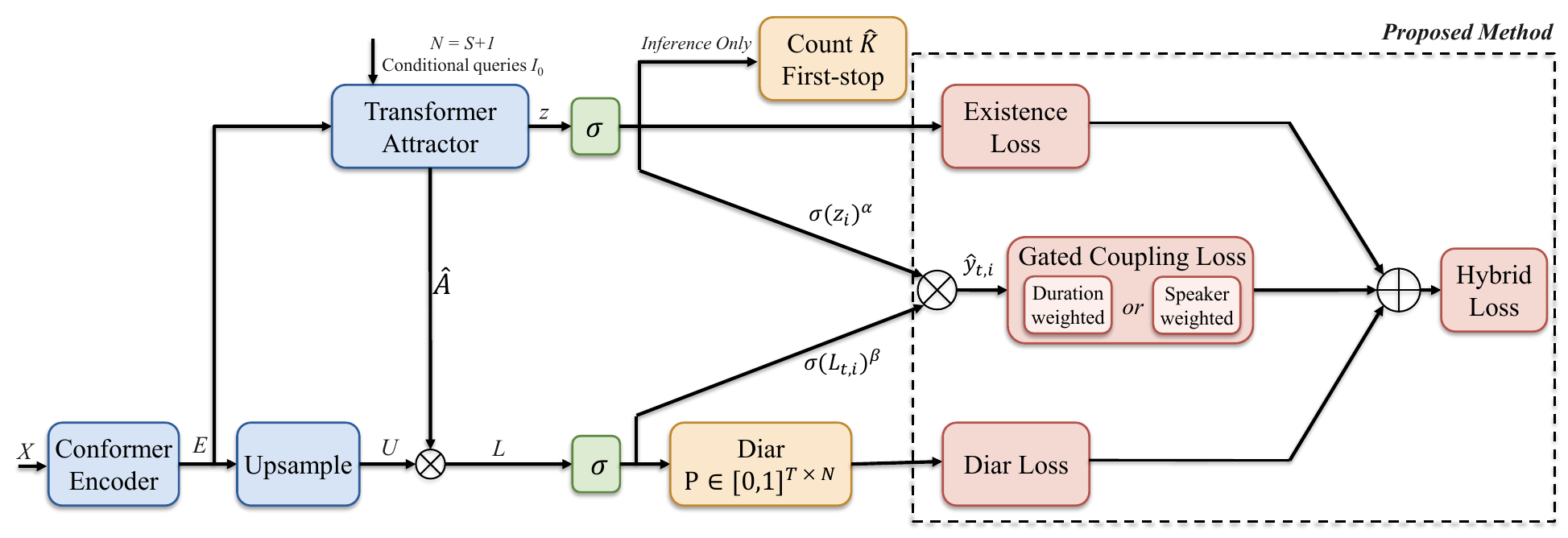}}
\caption{Overview of the proposed method on the EEND-TA backbone. The encoder feeds
a frame branch (upsampled frame embeddings $U$) and a Transformer attractor
(per-attractor existence logits $z$, $\ell_2$-normalized attractors $\hat{A}$), whose
dot product gives the diarization $P=\sigma(U\hat{A}^{\top})$. Our training-only
\emph{gated coupling loss} $\hat{y}_{t,i}=\sigma(z_i)^{\alpha}\sigma(L_{t,i})^{\beta}$
couples the existence and activity heads; it is added as a regularizer to the
Existence and Diar losses to form the Hybrid Loss, and its per-cell gated BCE is
reduced either duration- or speaker-weighted (Section~\ref{ssec:slotnorm}). The
count $\hat{K}$ (first-stop) is used at inference only.}
\label{fig:system}
\end{figure*}

Section~\ref{sec:analysis} argues that the under-counting in crowded recordings stems largely from how these models are trained. We focus on two properties of the standard training objective. It is \emph{decoupled}, supervising speaker existence and frame activity with separate terms. It is also \emph{duration-weighted}, so low-activity speakers are barely supervised. To address this, we introduce a training-only \emph{gated loss} that couples existence and activity. This gated loss can be reduced in two ways, \emph{duration-weighted} or \emph{speaker-weighted}. Fig.~\ref{fig:system} gives an overview.

\subsection{The EEND-TA backbone}
\label{ssec:backbone}

We recap our backbone, EEND-TA~\cite{samarakoon2023transformer}, to fix notation. Log-Mel features $X$ are encoded and temporally subsampled, then split
into a frame branch and an attractor branch,
\begin{equation}
\label{eq:backbone1}
\begin{gathered}
E = \operatorname{Encoder}(X) \in \mathbb{R}^{T' \times D}, \quad U = \operatorname{Upsample}(E) \in \mathbb{R}^{T \times D}, \\
(A, z) = \operatorname{TA}(E, c),
\end{gathered}
\end{equation}
with attractors $A \in \mathbb{R}^{N \times D}$ and per-attractor existence logits
$z \in \mathbb{R}^{N}$, where $N = S{+}1$ and $S$ is the maximum number of
emittable speakers. The encoder subsamples the input in time to length $T' < T$,
and $\operatorname{Upsample}$ restores the original frame rate; $\operatorname{TA}$ is
the Transformer attractor, conditioned on the conversation summary vector
$c$~\cite{broughton2023improving}. Diarization is a dot product of the upsampled frame
embeddings against the $\ell_2$-normalized attractors,
\begin{equation}
\label{eq:backbone2}
\hat{A} = A / \lVert A \rVert_2, \quad L = U \hat{A}^{\top} \in \mathbb{R}^{T \times N}, \quad P_{t,i} = \sigma(L_{t,i}).
\end{equation}
Here $\sigma(\cdot)$ is the sigmoid function. The attractors are emitted in descending existence order, so the speaker count is read
from the existence head by a \emph{first-stop} rule,
$\hat{K} = \min\{\, i : \sigma(z_i) \le \tau \,\}$, with indices from $i=0$ and threshold $\tau$. Since existence is non-increasing in $i$, attractors $0,\dots,\hat{K}-1$ are \emph{valid} (each represents a speaker) and the rest are \emph{empty}. If every attractor exceeds $\tau$, we set $\hat{K}=N$. EEND-TA trains two \emph{decoupled} terms, a
PIT activity BCE and an existence BCE,
\begin{equation}
\label{eq:diar_exist}
\begin{aligned}
\mathcal{L}_{\mathrm{diar}} &= \min_{\pi} \frac{1}{T S_b} \sum_{t,s} \operatorname{BCE}\!\left(L_{t,\pi(s)}, Y_{t,s}\right), \\
\mathcal{L}_{\mathrm{exist}} &= \frac{1}{N} \sum_{i} \operatorname{BCE}\!\left(z_i, \mathbf{1}[i < S_b]\right),
\end{aligned}
\end{equation}
combined with a fixed existence weight $\lambda_{\mathrm{exist}}$,
\begin{equation}
\label{eq:eendta}
\mathcal{L}_{\text{EEND-TA}} = \mathcal{L}_{\mathrm{diar}} + \lambda_{\mathrm{exist}}\, \mathcal{L}_{\mathrm{exist}},
\end{equation}
where the minimization over $\pi$ is solved by the per-recording Hungarian assignment. We denote its optimum $\pi^{*}$ and reuse it in the gated and reduction terms below. Here $S_b$ is the reference speaker count. These two terms are the decoupled, duration-weighted supervision that
Section~\ref{sec:analysis} links to the under-counting. The remaining
subsections develop a gated loss intended to counteract these two properties.

\subsection{Gated coupling loss}
\label{ssec:gated}

The decoupled terms permit an inconsistency that can contribute to mis-counting: an attractor
may be confidently ``exists'' yet silent, or active yet ``does-not-exist.'' The
gated loss, inspired by the set-prediction objective of
DETR~\cite{carion2020detr}, forces a single \emph{coupled} posterior. An attractor
counts as a speaker only if it both exists and is active,
\begin{equation}
\label{eq:gated_score}
\hat{y}_{t,i} = \sigma(z_i)^{\alpha}\, \sigma(L_{t,i})^{\beta},
\end{equation}
where the exponents $\alpha$ and $\beta$ set how decisively existence and activity
must each commit before an attractor is credited as a speaker. We hold $\beta$ fixed
and treat $\alpha$ as a tunable existence-gate sharpness. Their values are given in
Section~\ref{sec:exp}. We supervise this posterior with a matched-attractor term
over the cells assigned by the Hungarian match $\pi^{*}$, plus an empty-attractor
penalty that drives the existence of every empty attractor toward zero,
\begin{equation}
\label{eq:gated}
\mathcal{L}_{\mathrm{gated}} = \mathcal{L}_{\mathrm{match}} + w_{\varnothing}\, \frac{1}{|\varnothing|} \sum_{j \in \varnothing} \operatorname{softplus}(z_j),
\end{equation}
where $\varnothing$ is the set of empty attractors and $w_{\varnothing}$ weights the
empty-attractor penalty. The matched term
$\mathcal{L}_{\mathrm{match}}$ aggregates the per-cell gated BCE
$\operatorname{BCE}(\hat{y}_{t,\pi^{*}(s)}, Y_{t,s})$ over the matched set
$\mathcal{M}$ of (frame, speaker) pairs. \emph{How} this aggregation is performed is
the subject of Section~\ref{ssec:slotnorm}. On active frames ($Y{=}1$) the gated
product reduces to the decoupled form. The key effect is the conjunctive penalty on
$Y{=}0$ frames, where an attractor escapes penalty only by being low-existence
\emph{or} low-activity. The coupled product requires both factors to agree
before an attractor is credited. Using it \emph{alone}, in place of the decoupled
losses, over-constrains the model. We therefore keep the coupling as a regularizer
rather than a replacement. This is the role of the hybrid loss.

\subsection{Hybrid loss}
\label{ssec:hybrid}

Rather than replace the decoupled supervision, the hybrid loss \emph{adds} the
gated coupling as a regularizer while keeping the PIT and existence terms at full
strength,
\begin{equation}
\label{eq:hybrid}
\mathcal{L}_{\mathrm{hybrid}} = \mathcal{L}_{\mathrm{diar}} + \lambda_{\mathrm{exist}}\, \mathcal{L}_{\mathrm{exist}} + \lambda_{\mathrm{gated}}\, \mathcal{L}_{\mathrm{gated}}.
\end{equation}
Here $\lambda_{\mathrm{gated}}$ sets the strength of the coupling regularizer. The gated
term keeps an attractor's existence and activity
consistent, while the decoupled terms keep both heads directly supervised.
The coupling is thus imposed as a soft constraint on top of the original objective
rather than as a substitute for it.

\subsection{Duration-weighted vs.\ speaker-weighted reduction}
\label{ssec:slotnorm}

The matched term $\mathcal{L}_{\mathrm{match}}$ of Eq.~\eqref{eq:gated} must still
aggregate the per-cell gated BCE over the matched set $\mathcal{M}$, and this
choice decides \emph{which} speakers the supervision favors. We contrast two
reductions. Writing $\operatorname{BCE}_{t,s} \equiv
\operatorname{BCE}(\hat{y}_{t,\pi^{*}(s)}, Y_{t,s})$ for the per-cell term, the
\emph{duration-weighted} reduction pools every matched cell with
equal weight,
\begin{equation}
\label{eq:match_dur}
\mathcal{L}_{\mathrm{match}}^{\mathrm{dur}} = \frac{1}{|\mathcal{M}|} \sum_{(t,s) \in \mathcal{M}} \operatorname{BCE}_{t,s}.
\end{equation}
The active ($Y{=}1$) cells that push an attractor's existence upward are exactly
its speaker's active frames. This makes the positive supervision
proportional to its speaker's \emph{airtime}. A talkative speaker contributes many
active cells and dominates the gradient. The quiet, sparsely active speaker
of a crowded recording, the one most at risk of being dropped, receives the
least. The reduction therefore inherits the airtime bias of the
duration-weighted DER.

The coupling forces the model to factor frame-level activity into its existence
predictions, and the \emph{speaker-weighted} reduction instead gives each matched
attractor a single airtime-independent vote, balancing its active and silent frames
within the attractor before averaging over attractors. With $\mathcal{T}_s^{+}$ and
$\mathcal{T}_s^{-}$ the active and silent reference frames of speaker $s$ (matched to attractor $\pi^{*}(s)$),
\begin{equation}
\label{eq:match}
\begin{aligned}
\ell_s &= \frac{1}{2}\!\left( \frac{1}{|\mathcal{T}_s^{+}|} \sum_{t \in \mathcal{T}_s^{+}} \operatorname{BCE}_{t,s} + \frac{1}{|\mathcal{T}_s^{-}|} \sum_{t \in \mathcal{T}_s^{-}} \operatorname{BCE}_{t,s} \right), \\
\mathcal{L}_{\mathrm{match}}^{\mathrm{spk}} &= \frac{1}{S_b} \sum_{s} \ell_s.
\end{aligned}
\end{equation}
Every speaker, loud or quiet, now exerts the same existence pressure, so a rare,
low-airtime speaker is no longer drowned out by frequent, talkative ones. The two
reductions are two ways of computing the same gated loss. The
duration-weighted form mirrors the airtime emphasis of DER, whereas the
speaker-weighted form mirrors the speaker-weighted evaluation we care
about (JER and exact-count). This aligns the objective with the quantity we
set out to improve.

\section{Experiments}
\label{sec:exp}

\subsection{Datasets}

The pretraining data are obtained by simulating multi-speaker audio from the LibriSpeech Corpus \cite{librispeech} using the standard diarization mixture simulation algorithm \cite[Alg.~1]{fujita19eend}. In total, we generated 800,000 mixtures spanning 1 to 8 speakers. Following \cite{broughton2025pushing}, we adopt the speaker-count-dependent average silence intervals and apply the modification to the original mixture algorithm to reduce shared silence across all speakers. Specifically, the silence-interval parameter $\beta_{\mathrm{sil}}$ is $\{2, 2, 5, 9, 34, 54, 47, 50\}$ for mixtures that include 1–8 speakers. This resulted in 100,000 mixtures per speaker configuration and approximately 80,000 hours of data.

For fine-tuning, we use an aggregation of several widely used diarization corpora: AISHELL-4 \cite{aishell4}, AliMeeting \cite{alimeeting}, AMI-SDM (single distant microphone) and AMI-Mix (headset mixtures) \cite{AMI}, CallHome \cite{callhome}, DIHARD III \cite{ryant2020third}, MagicData-RAMC \cite{ramc}, and VoxConverse (v0.3) \cite{chung2020spot}. We adopt official data splits where available. For DIHARD III and VoxConverse, we treat the evaluation set as the test set and partition the training set into an 80\% / 20\% train / validation split. For datasets with multi-channel recordings, all channels are downmixed to a single channel. For evaluation we pool the test partitions of all corpora. Following Section~\ref{sec:analysis}, we report two speaker-count groups. The first is the full $\leq\!8$-speaker set ($811$ recordings). The second is the crowded $5$--$8$-speaker subset ($149$ recordings), where counting is hardest.

\subsection{Experimental Setup}

The input representation to the model is a sequence of 23-dimensional log-Mel filterbank features (25\,ms window, 10\,ms stride). The encoder subsamples them in time by a factor of 10, giving the length-$T'$ sequence of Eq.~\eqref{eq:backbone1}.

We employ a Conformer encoder with 6 blocks, a hidden size of 256, 4 attention heads, and a 1024-dimensional feed-forward layer per block. The Transformer attractor is a 3-layer Transformer decoder of the same width. It operates on the $N = S{+}1$ conditioned queries of the EEND-TA formulation~\cite{samarakoon2023transformer}. We set the maximum number of emittable speakers to $S = 8$, giving $N = 9$ attractors.


All models share a single backbone pretrained on the simulated LibriSpeech mixtures as mentioned in~\cite{broughton2025pushing}. The EEND-TA backbone is fine-tuned for $150$ epochs at batch size 8. We use Adam~\cite{kingma2014adam} at a constant learning rate of $1 \times 10^{-5}$ with gradient clipping. Recordings are cropped to $600$\,s, and chunk shuffling is applied~\cite{leung2021robust}.

For the loss coefficients we use an existence weight $\lambda_{\mathrm{exist}}{=}0.1$, a gated-loss weight $\lambda_{\mathrm{gated}}{=}0.5$, and an empty-attractor weight $w_{\varnothing}{=}0.1$ across all configurations. In the gated posterior of Eq.~\eqref{eq:gated_score} we fix $\beta{=}1$ throughout. We set $\alpha{=}1$ as the default and vary it only where noted in Section~\ref{sec:results}.

For every configuration the inference model uniformly averages the top-10 checkpoints ranked by validation DER, scored with the model's \emph{own} estimated speaker count. We rank by DER deliberately. It remains the metric most widely used in diarization, so selecting on it keeps our checkpoint choice biased toward aggregate accuracy rather than a better raw count. In practice, the counting gains we report come at comparable or better DER, not at its expense. All models use an attractor-existence threshold $\tau{=}0.5$ and a diarization threshold of $0.5$. All error rates are scored at a $0$\,s collar.

\section{Results}
\label{sec:results}

\subsection{Baselines}

Table~\ref{tab:baselines} compares our two candidate backbones under the standard diarization metrics (DER and JER) and the counting metrics of Section~\ref{sec:analysis}. The two backbones are the non-autoregressive EEND-TA and the autoregressive EEND-EDA, evaluated on both the full $\leq\!8$-speaker set and the crowded $5$--$8$ subset. The results confirm the trends of Fig.~\ref{figure_1} and Table~\ref{tab:count_bias}. Both systems perform well in aggregate but degrade sharply once recordings grow crowded. Exact-count accuracy drops from over $73\%$ to below $33\%$, and a strong negative count bias points to a tendency to under-predict speakers under dense overlap. Despite this shared weakness, EEND-TA outperforms EEND-EDA on every metric in both regimes. The gap is clearest on the crowded $5$--$8$ subset, where EEND-TA reaches $32.9\%$ exact-count accuracy against $20.8\%$ for EEND-EDA and also posts a lower DER.

\begin{table}[t]
\caption{Baseline diarization and speaker-counting performance of the non-autoregressive EEND-TA and autoregressive EEND-EDA.}
\begin{center}
\begin{tabularx}{\columnwidth}{|l|Y|Y|Y|Y|}
\hline
\textbf{Metric} & \multicolumn{2}{c|}{\textbf{EEND-TA (NAR)}} & \multicolumn{2}{c|}{\textbf{EEND-EDA (AR)}} \\
\cline{2-5}
 & \textbf{\textit{$\leq$8}} & \textbf{\textit{5--8}} & \textbf{\textit{$\leq$8}} & \textbf{\textit{5--8}} \\
\hline
DER (\%) $\downarrow$        & 13.24 & 18.29 & 14.18 & 20.35 \\
\hline
JER (\%) $\downarrow$        & 26.04 & 38.17 & 29.27 & 44.89 \\
\hline
Exact-count (\%) $\uparrow$  & 74.7  & 32.9  & 73.4  & 20.8  \\
\hline
 $\pm 1$ acc.\ (\%) $\uparrow$ & 91.9  & 62.4  & 90.1  & 59.1  \\
\hline
Count bias $\rightarrow\!0$  & $-0.17$ & $-1.03$ & $-0.19$ & $-1.13$ \\
\hline
\end{tabularx}
\label{tab:baselines}
\end{center}
\end{table}

\subsection{Training Objectives}
Table~\ref{tab:loss} compares EEND-TA trained under four loss objectives, each fine-tuned independently from the converged backbone. The first is the original decoupled losses (EEND-TA). The second replaces them with the gated coupling loss (+Gated). The last two keep the decoupled losses and add the gated term as a regularizer (the hybrid loss), under the duration-weighted and speaker-weighted reductions of its matched term. Replacing the decoupled losses with the gated loss alone hurts DER, which rises from $13.24$ to $13.92$ on the full set and from $18.29$ to $20.30$ on the crowded subset. It also worsens JER and yields no counting gain: crowded exact-count is unchanged at $32.9\%$ and $\pm 1$ accuracy drops to $59.7\%$. The coupling is thus useful only as a regularizer, not as a replacement. The hybrid loss returns DER to the baseline level ($13.18$ and $17.89$). But its matched term still pools over all frames, so the quiet, marginal speaker stays under-supervised. Crowded exact-count therefore slips to $28.2\%$. The speaker-weighted reduction closes this gap. Its airtime-independent per-attractor vote lifts crowded exact-count to $33.6\%$, raises $\pm 1$ accuracy from $62.4$ to $69.8\%$, roughly halves the count bias ($-1.03\!\rightarrow\!-0.62$), and improves JER by $2.5$ points. DER holds at the baseline level throughout ($17.20$ on the crowded subset, $13.14$ overall). The speaker-weighted reduction is thus the more effective of the two reductions, and the setting we build on next.

\begin{table}[t]
\caption{Speaker-counting and diarization performance of EEND-TA trained with different loss objectives. ``$+$\,Hybrid, dur.-wt.'' and ``$+$\,Hybrid, spk.-wt.'' are the hybrid loss under the duration-weighted and speaker-weighted reductions of its matched term (Sec.~\ref{ssec:slotnorm}).}
\begin{center}
\setlength{\tabcolsep}{3.2pt}
\footnotesize
\begin{tabularx}{\columnwidth}{|l|Y|Y|Y|Y|Y|}
\hline
\textbf{Model} & \textbf{DER} & \textbf{JER} & \textbf{Exact} & \textbf{$\pm 1$} & \textbf{Bias} \\
\hline
\multicolumn{6}{|c|}{\textit{all recordings ($\leq 8$ speakers)}} \\
\hline
EEND-TA               & 13.24 & 26.04 & 74.7 & 91.9 & $-0.17$ \\
$+$\,Gated            & 13.92 & 26.99 & 75.0 & 91.1 & $-0.22$ \\
$+$\,Hybrid, dur.-wt. & 13.18 & 26.50 & 73.7 & 91.4 & $-0.20$ \\
$+$\,Hybrid, spk.-wt. & 13.14 & 25.36 & 75.5 & 92.5 & $-0.09$ \\
\hline
\multicolumn{6}{|c|}{\textit{crowded subset ($5$--$8$ speakers)}} \\
\hline
EEND-TA               & 18.29 & 38.17 & 32.9 & 62.4 & $-1.03$ \\
$+$\,Gated            & 20.30 & 39.74 & 32.9 & 59.7 & $-1.13$ \\
$+$\,Hybrid, dur.-wt. & 17.89 & 38.44 & 28.2 & 60.4 & $-1.07$ \\
$+$\,Hybrid, spk.-wt. & 17.20 & 35.70 & 33.6 & 69.8 & $-0.62$ \\
\hline
\end{tabularx}
\label{tab:loss}
\end{center}
\end{table}

\subsection{Effect of the existence exponent $\alpha$}
Table~\ref{tab:alpha} sweeps the existence exponent $\alpha$ of the gated posterior (Eq.~\eqref{eq:gated_score}), which sets how strongly an attractor's existence gates its frame activity. We warm-start the sweep from the speaker-weighted hybrid model ($\alpha{=}1.0$) and report EEND-TA as the reference. Conceptually, a larger $\alpha$ super-linearly penalizes leaving a genuinely valid attractor's existence low. This strengthens the under-confident, low-airtime speakers that crowded recordings hinge on. The few speakers in sparse recordings already saturate the existence head, so the effect, and thus the gain, concentrates on the crowded tail. On the crowded subset, exact-count climbs from $33.6\%$ at $\alpha{=}1.0$ to $41.6\%$ at $\alpha{=}2.0$, with $\pm 1$ accuracy rising into the low-$70\%$ range and the count bias holding around $-0.6$. DER moves the same way: $\alpha{=}2.0$ posts the lowest DER in both panels ($12.98\%$ overall and $16.67\%$ on the crowded subset). $\alpha{=}2.0$ is the strongest operating point overall, so we adopt the speaker-weighted hybrid model at $\alpha{=}2.0$ as our \emph{final system} for the analyses that follow. The trend is not perfectly monotone. The $\alpha{=}1.7$ model is an outlier on crowded exact-count, and pushing beyond $\alpha{=}2.0$ to $\alpha{=}2.3$ regresses on both DER and crowded exact-count. Relative to the EEND-TA backbone, this training-only change lifts crowded exact-count from $32.9$ to $41.6\%$ and $\pm 1$ accuracy from $62.4$ to $72.5\%$. It roughly halves the count bias ($-1.03\!\rightarrow\!-0.60$) and cuts crowded JER from $38.2$ to $35.8\%$. All of this comes at no cost to DER ($18.3\!\rightarrow\!16.7\%$), the counting axis that duration-weighted DER scarcely registers.

\begin{table}[t]
\caption{DER, JER, and speaker-counting accuracy across the existence-exponent ($\alpha$) sweep of the speaker-weighted hybrid model; EEND-TA is the reference baseline.}
\label{tab:alpha}
\begin{center}
\setlength{\tabcolsep}{3.2pt}
\footnotesize
\begin{tabularx}{\columnwidth}{|l|Y|Y|Y|Y|Y|}
\hline
\textbf{Model} & \textbf{DER} & \textbf{JER} & \textbf{Exact} & \textbf{$\pm 1$} & \textbf{Bias} \\
\hline
\multicolumn{6}{|c|}{\textit{all recordings ($\leq 8$ speakers)}} \\
\hline
EEND-TA                     & 13.24 & 26.04 & 74.7 & 91.9 & $-0.17$ \\
$\alpha{=}1.0$ & 13.14 & 25.36 & 75.5 & 92.5 & $-0.09$ \\
$\alpha{=}1.3$ & 13.28 & 25.77 & 76.0 & 93.0 & $-0.10$ \\
$\alpha{=}1.5$ & 13.30 & 25.67 & 76.3 & 92.6 & $-0.05$ \\
$\alpha{=}1.7$ & 13.14 & 25.31 & 75.7 & 92.8 & $-0.06$ \\
$\alpha{=}2.0$ & 12.98 & 25.45 & 77.1 & 92.6 & $-0.08$ \\
$\alpha{=}2.3$ & 13.73 & 25.47 & 76.6 & 92.7 & $-0.05$ \\
\hline
\multicolumn{6}{|c|}{\textit{crowded subset ($5$--$8$ speakers)}} \\
\hline
EEND-TA                     & 18.29 & 38.17 & 32.9 & 62.4 & $-1.03$ \\
$\alpha{=}1.0$ & 17.20 & 35.70 & 33.6 & 69.8 & $-0.62$ \\
$\alpha{=}1.3$ & 17.19 & 36.51 & 37.6 & 73.2 & $-0.66$ \\
$\alpha{=}1.5$ & 17.90 & 36.35 & 38.9 & 73.8 & $-0.56$ \\
$\alpha{=}1.7$ & 17.55 & 36.06 & 29.5 & 73.8 & $-0.57$ \\
$\alpha{=}2.0$ & 16.67 & 35.76 & 41.6 & 72.5 & $-0.60$ \\
$\alpha{=}2.3$ & 18.15 & 36.10 & 35.6 & 72.5 & $-0.54$ \\
\hline
\end{tabularx}
\end{center}
\end{table}

\subsection{Missed-speaker and per-domain analysis}
Table~\ref{tab:missed_fate} traces \emph{where} the recovered speakers come from. We count a reference speaker as completely missed when its recall falls below $5\%$. For each missed speaker, we label its solo (non-overlapped) frames as absorbed into another speaker (Abs.), dropped to silence (Sil.), or absent because the speaker is only ever overlapped and hence unresolvable by a first-stop decoder (Ovl.). Of the $957$ reference speakers in the crowded subset, EEND-TA misses $209$ ($21.8\%$) entirely. Both speaker-weighted models cut this to about $160$ ($16.6$--$16.7\%$), recovering some fifty speakers. Almost all of that reduction comes from the silence category, which falls from $73$ to $33$--$37$. Absorptions decline only modestly ($120\!\rightarrow\!107$--$110$), and the genuinely unresolvable overlap-only count holds at $16$ across all three models. The speaker-weighted reduction thus rescues precisely the quiet, low-airtime speakers that duration-weighted supervision had let fall silent, not the overlap-only cases that no single first-stop pass can recover.

\begin{table}[t]
\caption{Fate of completely-missed reference speakers on the crowded $5$--$8$ subset: EEND-TA vs.\ the speaker-weighted hybrid model at $\alpha{=}1.0$ and $\alpha{=}2.0$.}
\label{tab:missed_fate}
\begin{center}
\setlength{\tabcolsep}{3pt}
\footnotesize
\begin{tabularx}{\columnwidth}{|l|Y|Y|Y|Y|}
\hline
\textbf{Model} & \textbf{Missed (\%)} & \textbf{Abs.} & \textbf{Sil.} & \textbf{Ovl.} \\
\hline
EEND-TA                     & $209\ (21.8)$ & 120 & 73 & 16 \\
$\alpha{=}1.0$ & $160\ (16.7)$ & 107 & 37 & 16 \\
$\alpha{=}2.0$ & $159\ (16.6)$ & 110 & 33 & 16 \\
\hline
\end{tabularx}
\end{center}
\end{table}

Table~\ref{tab:count_dom_ns58} breaks the crowded-subset counting down by domain
(only four corpora contain $5$--$8$-speaker recordings), comparing EEND-TA with our
speaker-weighted hybrid loss at $\alpha{=}2.0$. The gain holds broadly across domains.
It is largest on DIHARD-3 ($25\!\rightarrow\!42$ exact, $58\!\rightarrow\!75$
$\pm 1$, $-1.36\!\rightarrow\!-0.75$ bias) and AISHELL-4, which reaches $100\%$
$\pm 1$ accuracy at a near-zero $-0.10$ bias, and more modest on VoxConverse.
CallHome is the only exception: its exact-count drops ($25\!\rightarrow\!12$) while
$\pm 1$ accuracy is unchanged and the bias still edges toward zero. It lands
within one speaker as often as before, but pins the exact count less often. The
improvement therefore holds across most of the crowded corpora rather than resting
on a single domain, with CallHome the one exception, which we examine next.

CallHome behaves this way because our mechanism can recover only one kind of
missed speaker. Raising a speaker's existence confidence above the counting
threshold lifts back a \emph{silenced} speaker, one dropped below threshold but
still localizable. This is the silence-fate reduction behind the gains of
Table~\ref{tab:missed_fate}. An \emph{absorbed} speaker, by contrast, has already
been folded onto a higher-airtime neighbour's slot, a merge that existence
pressure cannot reverse. On CallHome the crowded missed speakers are almost
entirely absorbed rather than silenced, so the objective only converts its few
silenced cases into absorbed ones and the missed-speaker count stays flat. With
nothing to recover, the added existence pressure instead surfaces as false
alarms. CallHome's crowded false-alarm rate rises while its missed speech
falls, which accounts for the small DER regression. The exact-count dip itself
turns on a single one of the eight crowded calls, so with so few recordings it
should be read cautiously. CallHome thus pays the method's cost without its benefit.

On the sparse regime ($1$--$4$ speakers) counting is already near-saturated: exact-count at
$84$--$85\%$, $\pm 1$ accuracy near $98\%$, and essentially zero bias. The
speaker-weighted objective neither helps nor hurts it (the fixed-roster meeting
domains AMI, AliMeeting, and MagicData are already at $100\%$ exact), so we
omit its per-domain breakdown.

\begin{table}[t]
\caption{Domain-level speaker counting on crowded recordings ($5$--$8$ speakers): EEND-TA vs.\ the speaker-weighted $\alpha{=}2.0$ model.}
\label{tab:count_dom_ns58}
\begin{center}
\setlength{\tabcolsep}{3pt}
\footnotesize
\begin{tabularx}{\columnwidth}{|l|YYY|YYY|}
\hline
\textbf{Domain} & \multicolumn{3}{c|}{\textbf{EEND-TA}} & \multicolumn{3}{c|}{\textbf{$\alpha{=}2.0$}} \\
\cline{2-7}
 & \textbf{Ex.} & \textbf{$\pm1$} & \textbf{Bias} & \textbf{Ex.} & \textbf{$\pm1$} & \textbf{Bias} \\
\hline
DIHARD-3    & 25 & 58 & $-1.36$ & 42 & 75 & $-0.75$ \\
CallHome    & 25 & 62 & $-1.12$ & 12 & 62 & $-1.00$ \\
VoxConverse & 29 & 59 & $-0.99$ & 38 & 66 & $-0.61$ \\
AISHELL-4   & 65 & 85 & $-0.55$ & 70 & 100 & $-0.10$ \\
\hline
\textbf{All} & 33 & 62 & $-1.03$ & 42 & 72 & $-0.60$ \\
\hline
\end{tabularx}
\end{center}
\end{table}

\section{Conclusions}
\label{sec:conclusion}

We revisited speaker counting in end-to-end neural diarization. We showed that both autoregressive EEND-EDA and non-autoregressive EEND-TA
systematically under-count as recordings grow crowded. The duration-weighted DER
conceals this failure. Dominated by talkative speakers, it stays low even as quiet
ones vanish, with the count bias reaching nearly two missing speakers at eight
participants. Measuring counting directly with JER, exact-count, $\pm 1$ accuracy,
and signed bias makes this collapse visible. We argue this collapse stems largely
from how the models are trained. Duration-weighted supervision appears to give
low-activity speakers little gradient, so that they tend to be neither counted nor
assigned frames. Our remedy is purely a training-time change to EEND-TA that leaves
the backbone and inference untouched. It adds a detection-inspired gated loss that couples existence with activity,
applied as a hybrid regularizer. A per-attractor speaker-weighted reduction gives
every speaker one airtime-independent vote. Sharpening the existence gate pushes counting further. On
pooled real-world data the method raises crowded-subset exact-count from $32.9$ to
$41.6\%$, lifts $\pm 1$ accuracy to $72.5\%$, and roughly halves the count bias. This
comes at no cost to DER.

\newpage
\section{AI-Generated Content Disclosure}
We used Claude for coding, figure design and checking the narrative and the language structures of the paper. 
All technical content was reviewed and verified by the authors, who take full responsibility for the final work.
\bibliographystyle{IEEEtran}
\bibliography{refs}

\begin{thebibliography}{10}
\providecommand{\url}[1]{#1}
\csname url@samestyle\endcsname
\providecommand{\newblock}{\relax}
\providecommand{\bibinfo}[2]{#2}
\providecommand{\BIBentrySTDinterwordspacing}{\spaceskip=0pt\relax}
\providecommand{\BIBentryALTinterwordstretchfactor}{4}
\providecommand{\BIBentryALTinterwordspacing}{\spaceskip=\fontdimen2\font plus
\BIBentryALTinterwordstretchfactor\fontdimen3\font minus
  \fontdimen4\font\relax}
\providecommand{\BIBforeignlanguage}[2]{{%
\expandafter\ifx\csname l@#1\endcsname\relax
\typeout{** WARNING: IEEEtran.bst: No hyphenation pattern has been}%
\typeout{** loaded for the language `#1'. Using the pattern for}%
\typeout{** the default language instead.}%
\else
\language=\csname l@#1\endcsname
\fi
#2}}
\providecommand{\BIBdecl}{\relax}
\BIBdecl

\bibitem{ryant2020third}
N.~Ryant, P.~Singh, V.~Krishnamohan, R.~Varma, K.~Church, C.~Cieri, J.~Du,
  S.~Ganapathy, and M.~Liberman, ``{The Third DIHARD Diarization Challenge},''
  in \emph{Proc. Interspeech 2021}, 2021, pp. 3570--3574.

\bibitem{chung2020spot}
J.~S. Chung, J.~Huh, A.~Nagrani, T.~Afouras, and A.~Zisserman, ``{Spot the
  Conversation: Speaker Diarisation in the Wild},'' in \emph{Proc. Interspeech
  2020}, 2020, pp. 299--303.

\bibitem{kinoshita21_interspeech}
K.~Kinoshita, M.~Delcroix, and N.~Tawara, ``{Advances in Integration of
  End-to-End Neural and Clustering-Based Diarization for Real Conversational
  Speech},'' in \emph{Proc. Interspeech 2021}, 2021, pp. 3565--3569.

\bibitem{anguera2012speaker}
X.~Anguera, S.~Bozonnet, N.~Evans, C.~Fredouille, G.~Friedland, and O.~Vinyals,
  ``Speaker diarization: A review of recent research,'' \emph{IEEE Transactions
  on audio, speech, and language processing}, vol.~20, no.~2, pp. 356--370,
  2012.

\bibitem{park2022review}
T.~J. Park, N.~Kanda, D.~Dimitriadis, K.~J. Han, S.~Watanabe, and S.~Narayanan,
  ``A review of speaker diarization: Recent advances with deep learning,''
  \emph{Computer Speech \& Language}, vol.~72, p. 101317, 2022.

\bibitem{diez2018but}
M.~Diez, F.~Landini, L.~Burget, J.~Rohdin, A.~Silnova, K.~Zmol{\'\i}kov{\'a},
  O.~Novotn{\`y}, K.~Vesel{\`y}, O.~Glembek, O.~Plchot \emph{et~al.}, ``But
  system for dihard speech diarization challenge 2018.'' in \emph{Proc.
  Interspeech 2018}, 2018, pp. 2798--2802.

\bibitem{fujita2019end}
Y.~Fujita, N.~Kanda, S.~Horiguchi, Y.~Xue, K.~Nagamatsu, and S.~Watanabe,
  ``End-to-end neural speaker diarization with self-attention,'' in \emph{2019
  IEEE Automatic Speech Recognition and Understanding Workshop (ASRU)}, 2019,
  pp. 296--303.

\bibitem{yu2017permutation}
D.~Yu, M.~Kolb{\ae}k, Z.-H. Tan, and J.~Jensen, ``Permutation invariant
  training of deep models for speaker-independent multi-talker speech
  separation,'' in \emph{Proc. ICASSP 2017}, 2017, pp. 241--245.

\bibitem{fujita19eend}
Y.~Fujita, N.~Kanda, S.~Horiguchi, K.~Nagamatsu, and S.~Watanabe, ``{End-to-End
  Neural Speaker Diarization with Permutation-Free Objectives},'' in
  \emph{Proc. Interspeech 2019}, 2019, pp. 4300--4304.

\bibitem{horiguchi20EENDEDA}
S.~Horiguchi, Y.~Fujita, S.~Watanabe, Y.~Xue, and K.~Nagamatsu, ``{End-to-End
  Speaker Diarization for an Unknown Number of Speakers with Encoder-Decoder
  Based Attractors},'' in \emph{Proc. Interspeech 2020}, 2020, pp. 269--273.

\bibitem{horiguchi22EENDEDA}
S.~Horiguchi, Y.~Fujita, S.~Watanabe, Y.~Xue, and P.~Garcia, ``{Encoder-Decoder
  Based Attractors for End-to-End Neural Diarization},'' \emph{{IEEE}/{ACM}
  Transactions on Audio, Speech, and Language Processing}, vol.~30, pp.
  1493--1507, 2022.

\bibitem{samarakoon2023transformer}
L.~Samarakoon, S.~J. Broughton, M.~H{\"a}rk{\"o}nen, and I.~Fung, ``Transformer
  attractors for robust and efficient end-to-end neural diarization,'' in
  \emph{2023 IEEE Automatic Speech Recognition and Understanding Workshop
  (ASRU)}, 2023, pp. 1--8.

\bibitem{transformer}
A.~Vaswani, N.~Shazeer, N.~Parmar, J.~Uszkoreit, L.~Jones, A.~N. Gomez,
  L.~Kaiser, and I.~Polosukhin, ``{Attention is All You Need},'' in
  \emph{Advances in Neural Information Processing Systems (NeurIPS)}, 2017, pp.
  6000--6010.

\bibitem{gulati2020conformer}
A.~Gulati, J.~Qin, C.-C. Chiu, N.~Parmar, Y.~Zhang, J.~Yu, W.~Han, S.~Wang,
  Z.~Zhang, Y.~Wu, and R.~Pang, ``{Conformer: Convolution-augmented Transformer
  for Speech Recognition},'' in \emph{Proc. Interspeech 2020}, 2020, pp.
  5036--5040.

\bibitem{broughton2023improving}
S.~J. Broughton and L.~Samarakoon, ``{Improving End-to-End Neural Diarization
  Using Conversational Summary Representations},'' in \emph{Proc. Interspeech
  2023}, 2023, pp. 3157--3161.

\bibitem{rybicka2022end}
M.~Rybicka, J.~Villalba, N.~Dehak, and K.~Kowalczyk, ``End-to-end neural
  speaker diarization with an iterative refinement of non-autoregressive
  attention-based attractors,'' in \emph{Proc. Interspeech 2022}, 2022, pp.
  5090--5094.

\bibitem{fujita2023intermediate}
Y.~Fujita, T.~Komatsu, R.~Scheibler, Y.~Kida, and T.~Ogawa, ``Neural
  diarization with non-autoregressive intermediate attractors,'' in \emph{Proc.
  ICASSP 2023}, 2023, pp. 1--5.

\bibitem{horiguchi21global_local}
S.~Horiguchi, S.~Watanabe, P.~García, Y.~Xue, Y.~Takashima, and Y.~Kawaguchi,
  ``Towards neural diarization for unlimited numbers of speakers using global
  and local attractors,'' in \emph{2021 IEEE Automatic Speech Recognition and
  Understanding Workshop (ASRU)}, 2021, pp. 98--105.

\bibitem{kanda2022transcribe}
N.~Kanda, X.~Xiao, Y.~Gaur, X.~Wang, Z.~Meng, Z.~Chen, and T.~Yoshioka,
  ``Transcribe-to-diarize: Neural speaker diarization for unlimited number of
  speakers using end-to-end speaker-attributed asr,'' in \emph{Proc. ICASSP
  2022}, 2022, pp. 8082--8086.

\bibitem{ryant2020thirdeval}
N.~Ryant, K.~Church, C.~Cieri, J.~Du, S.~Ganapathy, and M.~Liberman, ``Third
  {DIHARD} challenge evaluation plan,'' \emph{arXiv preprint arXiv:2006.05815},
  2020.

\bibitem{ryant2019second}
N.~Ryant, K.~Church, C.~Cieri, A.~Cristia, J.~Du, S.~Ganapathy, and
  M.~Liberman, ``The second dihard diarization challenge: Dataset, task, and
  baselines,'' in \emph{Proc. Interspeech 2019}, 2019, pp. 978--982.

\bibitem{yu2022auxiliary}
Y.~Yu, D.~Park, and H.~K. Kim, ``Auxiliary loss of transformer with residual
  connection for end-to-end speaker diarization,'' in \emph{Proc. ICASSP 2022},
  2022, pp. 8377--8381.

\bibitem{samarakoon2025variance}
L.~Samarakoon, S.~J. Broughton, and I.~Fung, ``Variance-covariance
  regularization for improved end-to-end diarization,'' in \emph{Proc. ICASSP
  2025}, 2025, pp. 1--5.

\bibitem{carion2020detr}
N.~Carion, F.~Massa, G.~Synnaeve, N.~Usunier, A.~Kirillov, and S.~Zagoruyko,
  ``End-to-end object detection with transformers,'' in \emph{European
  Conference on Computer Vision (ECCV)}, 2020, pp. 213--229.

\bibitem{harkonen2024eend}
M.~H{\"a}rk{\"o}nen, S.~J. Broughton, and L.~Samarakoon, ``Eend-m2f:
  Masked-attention mask transformers for speaker diarization,'' \emph{arXiv
  preprint arXiv:2401.12600}, 2024.

\bibitem{librispeech}
V.~Panayotov, G.~Chen, D.~Povey, and S.~Khudanpur, ``{Librispeech: An ASR
  Corpus Based on Public Domain Audio Books},'' in \emph{Proc. ICASSP 2015},
  2015, pp. 5206--5210.

\bibitem{broughton2025pushing}
S.~J. Broughton and L.~Samarakoon, ``Pushing the limits of end-to-end
  diarization,'' in \emph{Proc. Interspeech 2025}, 2025, pp. 5218--5222.

\bibitem{aishell4}
Y.~Fu, L.~Cheng, S.~Lv, Y.~Jv, Y.~Kong, Z.~Chen, Y.~Hu, L.~Xie, J.~Wu, H.~Bu,
  X.~Xu, J.~Du, and J.~Chen, ``{AISHELL-4: An Open Source Dataset for Speech
  Enhancement, Separation, Recognition and Speaker Diarization in Conference
  Scenario},'' in \emph{Proc. Interspeech 2021}, 2021, pp. 3665--3669.

\bibitem{alimeeting}
F.~Yu, S.~Zhang, Y.~Fu, L.~Xie, S.~Zheng, Z.~Du, W.~Huang, P.~Guo, Z.~Yan,
  B.~Ma, X.~Xu, and H.~Bu, ``M2{M}e{T}: The {ICASSP} 2022 multi-channel
  multi-party meeting transcription challenge,'' in \emph{Proc. ICASSP 2022},
  2022, pp. 6167--6171.

\bibitem{AMI}
W.~Kraaij, T.~Hain, M.~Lincoln, and W.~Post, ``The ami meeting corpus,'' in
  \emph{Proc. International Conference on Methods and Techniques in Behavioral
  Research}, 2005.

\bibitem{callhome}
NIST, ``The 2000 {NIST} speaker recognition evaluation plan.'' National
  Institute of Standards and Technology (NIST), Tech. Rep., 2009.

\bibitem{ramc}
Z.~Yang, Y.~Chen, L.~Luo, R.~Yang, L.~Ye, G.~Cheng, J.~Xu, Y.~Jin, Q.~Zhang,
  P.~Zhang, L.~Xie, and Y.~Yan, ``{Open Source MagicData-RAMC: A Rich Annotated
  Mandarin Conversational(RAMC) Speech Dataset},'' in \emph{Proc. Interspeech
  2022}, 2022, pp. 1736--1740.

\bibitem{kingma2014adam}
D.~P. Kingma and J.~Ba, ``{Adam: A Method for Stochastic Optimization},'' in
  \emph{3rd International Conference on Learning Representations (ICLR)}, 2015.

\bibitem{leung2021robust}
T.-Y. Leung and L.~Samarakoon, ``{Robust End-to-End Speaker Diarization with
  Conformer and Additive Margin Penalty},'' in \emph{Proc. Interspeech 2021},
  2021, pp. 3575--3579.

\end{thebibliography}
\end{document}